\documentstyle[11pt,newpasp,twoside,epsf]{article}

\def\bvr{\hbox{$BV\!R$}}
\def\fdeg{\hbox{$.\mkern-6mu^\circ$}}
\def\solar{\mbox{$_{\normalsize\odot}$}}
\def\alp{\mbox{$\alpha$}}
\def\ox{\hbox{$\otimes$}}
\def\plotsmall#1{\centering \leavevmode
\epsfxsize=0.55\textwidth \epsfbox{#1}}

\begin{document}

\title{Stellar Associations and their Field East of LMC 4 in the Large 
Magellanic Cloud}

\author{D. Gouliermis and K. S. de Boer}
\affil{Sternwarte der Universit\"{a}t Bonn, Auf dem H\"{u}gel
        71, D-53121 Bonn, Germany}
\author{S. C. Keller}
\affil{Institute of Geophysics and Planetary Physics, Lawerence
Livermore Nat. Lab., L-413, P.O. Box 808, Livermore, CA 94550,
U.S.A}
\author{M. Kontizas}
\affil{Department of Astrophysics Astronomy \& Mechanics, Faculty
of Physics, University of Athens, GR-157 83 Athens, Greece}
\author{E. Kontizas}
\affil{Institute for Astronomy and Astrophysics, National
Observatory of Athens, P.O. Box 20048, GR-118 10 Athens, Greece}

\begin{abstract}
We report about the stellar content and the luminosity and mass functions
of three stellar associations and their field located on the north-east
edge of the super-bubble LMC 4 in the Large Magellanic Cloud.
\end{abstract}

\section{Introduction}

OB stellar associations represent the younger stellar systems in a galaxy.  
The Large Magellanic Cloud (LMC) is characterised by a large number of
such systems, as well as of giant and super-giant shells detected in {\sc
Hi}. The edges of super-giant shells are in many cases the loci of recent
star formation. In order to investigate the population of young stellar
systems of the LMC, Gouliermis et al. (2000a) developed an objective
technique for the detection of low concentrated young stellar groups.  
The detected systems are classified according to their stellar densities
(in M{\solar} pc$^{-3}$) into three categories, namely `unbound',
`intermediate' and `bound' systems. It was found that the unbound systems
are the stellar associations of the galaxy, while most of the intermediate
($\sim$ 70\%) are probably stellar associations or open clusters.

The application of this method on a digitised 1.2m UK Schmidt Telescope
Plate in the U band resulted in a Catalog of young stellar systems in the
central (6\fdeg5 $\times$ 6\fdeg5) region of the LMC (Gouliermis et al.
2000b). It was found that the systems show to be aligned in filamentary
structures (or arcs). A comparison of this survey with the newly detected
giant and super-giant shells by Kim et al. (1999)  showed that some of
these arcs are related to the edges of the shells (Figure 1). We report
the results of an investigation of stellar associations located on the
north-east edge of one of the most interesting super-giant shells in this
galaxy, LMC 4.

\begin{figure}[t]
\plotsmall{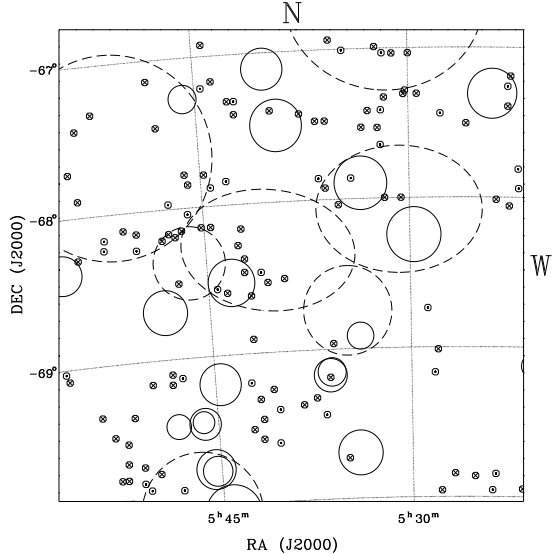}
\caption{The NE part of the Survey of young stellar systems 
in the central area of the LMC by Gouliermis et al. (2000b). The {\sun} 
symbols represent the loci of the detected `unbound' stellar systems, 
while the {\ox} symbols represent the `intermediate' ones. The areas of  
the giant (solid lines) and super-giant (dashed lines) shells 
as were identified by Kim et al. (1999) have been overplotted. Only a 
small portion of LMC 4 area, shown on the NW part of the figure, is 
covered by the survey.}
\end{figure}

\section{Stellar Content of the Observed Systems and the Field}

We performed {\bvr} photometry in an area of 20\farcm5 $\times$ 20\farcm5
to investigate the stellar content of three stellar associations: LH 91 \&
LH 95 (Lucke \& Hodge 1970) and LH 91-I (Kontizas et al. 1994) and their
field, situated to the east of LMC 4. Our observations include H{\alp}
measurements to identify the Be population of the region. We found that Be
stars exist in all the associations. In LH 95 we verified that the
{\sc Hii} emission is strongly related to Be stars located in the very 
centre of the system. We estimated the reddening and the age of the 
systems based on isochrone fitting. The reddening was found to vary 
between $E(B-V) \simeq 0.15$ and 0.20 mag. All systems were found to be 
younger than 10 Myr, while the field is older than $\sim 80$ Myr. Both LH 
91-I and LH 95 were identified as distinct stellar groups using star 
counts, while LH 91 does not show to represent any specific stellar 
concentration.

\begin{figure}[t]
\plotone{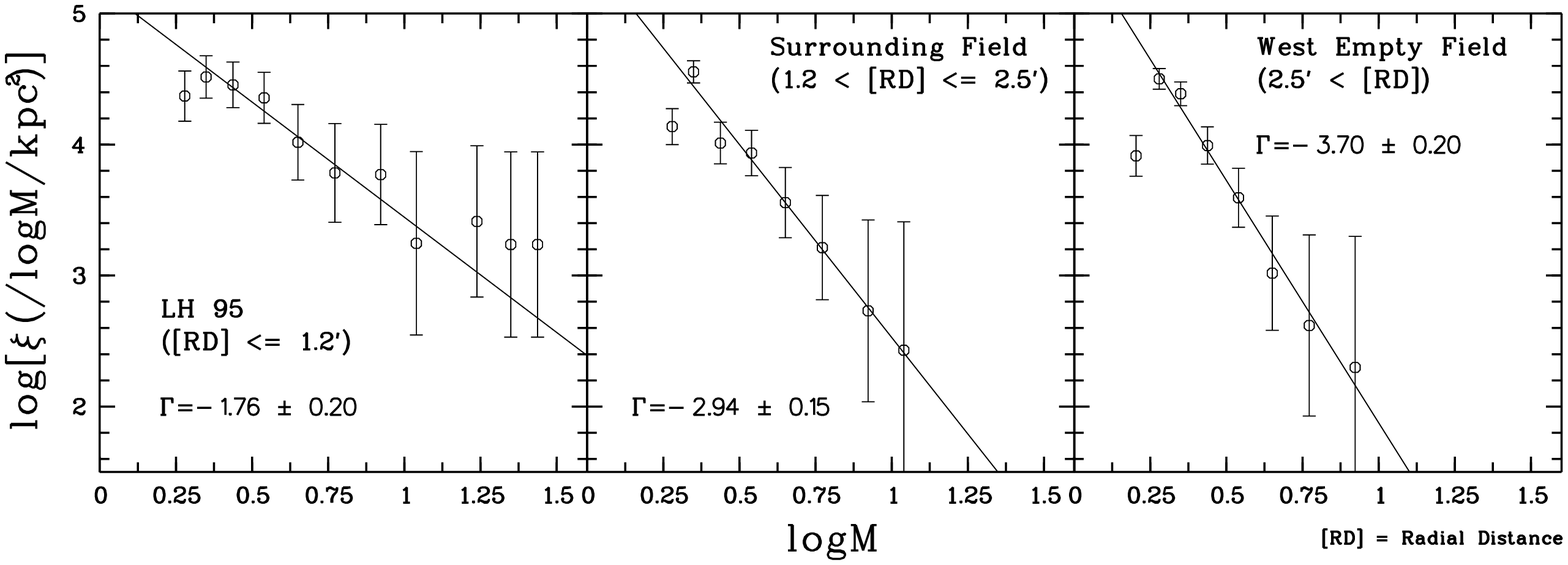}
\caption{Radial gradient of the MF slope outwards the centre of LH 
95, within 1\farcm2 (left), for radial distances 1\farcm2 $<r<$ 2\farcm5 
(middle) and for a more distant empty area toward the west, where there is 
no contamination of stars from other systems (right).} 
\end{figure}

\subsection{Luminosity and Mass Functions}

We constructed the luminosity functions (LFs) of the MS populations of the
systems and the field using the $B$ magnitudes. The slopes $s$ of the
completeness corrected and field subtracted LFs ($\log{N} \propto s \cdot
B_{\rm bin}$) of LH 91-I and LH 95 were found to be around $0.16 \pm
0.06$, while the one of LH 91 is steeper and similar to the field's ($s
\simeq 0.51 \pm 0.08$) for the same magnitude range of 16.5 $\la B \la$
19.5 mag.

For the construction of the mass functions (MFs) we performed counts of
stars in mass intervals. We corrected the counted numbers for
incompleteness and we normalized them to a surface of 1 kpc$^{2}$. A
Salpeter IMF corresponds to slope $\Gamma = -1.35$. It was found that the
systems have MF slopes in the range between $\Gamma \simeq -1.6 \pm 0.3$
(field subtracted MF of LH 95)  and $-2.2 \pm 0.4$ (LH 91), much more
shallow than the MF slope of the field, which varies from $\Gamma = -3.7$
up to $-4.9$.

In Figure 2 the most interesting case of LH 95 is presented, where a
radial dependence of the MF slope was observed, possibly an indication
that stars escape from the system contributing to the enrichement of the
field population. We verified that LH 95 is not an extended system
(further than $\sim$ 1\farcm5), and probably is being disrupted 
(Gouliermis et al. 2000c).

\end{document}